\documentclass[12pt, preprint, epsfig]{aastex6}
\usepackage{graphicx}
\usepackage{color}
\usepackage{epstopdf}
\begin{document}
\title{First Observational Signature of Rotational Deceleration in a
  Massive, Intermediate-age Star Cluster in the Magellanic Clouds}
\author{
Xiaohan Wu\altaffilmark{1,2},
Chengyuan Li\altaffilmark{2,3,4,5},
Richard de Grijs\altaffilmark{2,6}, and
Licai Deng\altaffilmark{3}
}

\altaffiltext{1} {School of Physics, Peking University, Yi He Yuan Lu
  5, Hai Dian District, Beijing 100871, China} 
\altaffiltext{2} {Kavli Institute for Astronomy \& Astrophysics and
  Department of Astronomy, Peking University, Yi He Yuan Lu 5, Hai
  Dian District, Beijing 100871, China; grijs@pku.edu.cn}
\altaffiltext{3} {Key Laboratory for Optical Astronomy, National
  Astronomical Observatories, Chinese Academy of Sciences, 20A Datun
  Road, Chaoyang District, Beijing 100012, China}
\altaffiltext{4} {Department of Physics and Astronomy, Macquarie
  University, Sydney, NSW 2109, Australia}
\altaffiltext{5} {Purple Mountain Observatory, Chinese Academy of
  Sciences, Nanjing 210008, China}
\altaffiltext{6} {International Space Science Institute--Beijing, 1
  Nanertiao, Zhongguancun, Hai Dian District, Beijing 100190, China}

\begin{abstract}
While the extended main-sequence turn-offs (eMSTOs) found in almost
all 1--2 Gyr-old star clusters in the Magellanic Clouds are often
explained by postulating extended star-formation histories, the tight
subgiant branches (SGBs) seen in some clusters challenge this popular
scenario. Puzzlingly, the SGB of the eMSTO cluster NGC 419 is
significantly broader at bluer than at redder colors. We carefully
assess and confirm the reality of this observational trend. If we
would assume that the widths of the features in color--magnitude space
were entirely owing to a range in stellar ages, the star-formation
histories of the eMSTO stars and the blue SGB region would be
significantly more prolonged than that of the red part of the
SGB. This cannot be explained by assuming an internal age spread. We
show that rotational deceleration of a population of rapidly rotating
stars, a currently hotly debated alternative scenario, naturally
explains the observed trend along the SGB. Our analysis shows that a
`converging' SGB could be produced if the cluster is mostly composed
of rapidly rotating stars that slow down over time owing to the
conservation of angular momentum during their evolutionary expansion
from main-sequence turn-off stars to red giants.
\end{abstract}


\section{Introduction}

The once common perception that the member stars of most star clusters
originate from `simple stellar populations' (SSPs) is increasingly
challenged by discoveries of extended main-sequence turn-offs (eMSTOs)
in the color--magnitude diagrams (CMDs) of intermediate-age, 1--2
Gyr-old massive star clusters in the Magellanic Clouds
\citep[e.g.,][]{Mack07,Mack08,Milo09,Rube10,Gira13,Li14a}. If solely
interpreted in terms of stellar age distributions, such eMSTOs may
imply age spreads greater than 300 Myr
\citep[e.g.,][]{Rube10,Rube11,Goud14}. However, this strongly
contradicts our current understanding of the evolution of SSP-like
star clusters, whose maximum age spreads are expected to reach only
1--3 Myr \citep{Long14}.

\cite{Li14b} were the first to focus on the subgiant-branch (SGB)
morphology of an eMSTO cluster, NGC 1651. Its tight SGB is
inconsistent with the presence of a significant age
spread. Simultaneously, \cite{BN15} discovered that the SGB and red
clump morphologies of two other intermediate-age star clusters, NGC
1806 and NGC 1846, also favor SSP scenarios. Several authors have
shown that the presence of a range of stellar rotation rates in a
single-age stellar population can produce the observed eMSTO features
\citep[e.g.,][]{Bast09,Yang13,Bran15,Nied15}. In addition, a
sufficiently low mixing efficiency could generate a tight SGB,
resembling the observed SGB morphologies \citep{Yang13}. Recently, the
stellar rotation scenario received further support from \cite{Milo16},
who discovered that the split main sequences in the young Large
Magellanic Cloud cluster NGC 1755 could best be explained by stellar
rotation, and not by adopting an age spread.

Here we report the discovery of a `converging' SGB in the Small
Magellanic Cloud eMSTO star cluster NGC 419: its SGB is significantly
broader on the blue than on the red side. The apparently prolonged
star-formation histories (SFHs) implied by the eMSTO stars and the
blue region of the SGB contradict that derived for the red SGB
region. We conclude that the observed SGB morphology can be explained
by rotational deceleration owing to the conservation of angular
momentum during the evolutionary expansion of SGB stars.

\section{Data Reduction}

We used archival images from {\sl Hubble Space Telescope} ({\sl HST})
program GO-10396 (PI: J. S. Gallagher), which employed the Advanced
Camera for Surveys (ACS)/Wide Field Channel (WFC). The individual
images in the F555W and F814W filters cover $200'' \times 200''$,
offset by $37''$ from the cluster center. The F555W and F814W filters
closely resemble the Johnson--Cousins $V$ and $I$ bands,
respectively. We downloaded four images with total exposure times of
1984 s and 1896 s in the F555W ($V$) and F814W ($I$) filters,
respectively, as well as two images with total exposure times of 40 s
and 20 s in these filters. To facilitate field-star decontamination,
we also retrieved images of a nearby field region at some $230''$ from
the cluster center, observed as part of the same {\it HST} program
with the same instrumental setup. The latter observations, covering
the same total area on the sky, include three images with total
exposure times of 1617 s and 1512 s and one image each with exposure
times of 50 s and 48 s in the $V$ and $I$ bands, respectively.

We performed point-spread-function (PSF) photometry using both {\sc
  iraf/daophot} \citep{Davis94} and {\sc dolphot}'s ACS module
\citep{Dolphin11,Dolphin13}. We adopted the {\sc dolphot} stellar
catalog for our analysis and used the {\sc iraf/daophot} photometry
for comparison and consistency checks (see Appendix). We determined
the cluster center by fitting Gaussian functions to the cluster's
number-density profiles along the right ascension ($\alpha_{\rm
  J2000}$) and declination ($\delta_{\rm J2000}$) axes: $\alpha_{\rm
  J2000}=01^{\rm h}08^{\rm m}17.02^{\rm s}$, $\delta_{\rm
  J2000}=-72^{\circ}53'3.12''$. This center position compares very
well with previous
determinations.\footnote{http://ned.ipac.caltech.edu/cgi-bin/objsearch?objname=ngc+419,
  http://simbad.u-strasbg.fr/simbad/sim-basic?Ident=ngc+419.} We
obtained the cluster radius, $R$, based on \citet[][EFF]{Elson1987}
profile fits (appropriate for non-tidally truncated clusters) to the
completeness-corrected projected radial stellar number density,
adopting for the cluster radius five times its core radius, i.e.,
$R=75''$ (see Appendix).

Next, we performed statistical field-star decontamination. We divided
the cluster and field CMDs into 15 bins along the color axis, $-1.0
\le (V-I) \le 3.0$ mag, and 28 bins along the magnitude axis, $16.0
\le V \le 27.0$ mag. We measured the number of field stars in each bin
in the field CMD and randomly removed the appropriate number---scaled
by the ratio of the area within $R = 75''$ to the total area covered
by the field CMD---from the corresponding bin of the cluster CMD. If
the number of stars that had to be removed was larger than the number
of stars present in a given bin, all stars were removed. To ascertain
the reliability of this approach, we also employed a method similar to
that of \cite{Cabr16}, which showed that the SGB region is
characterized by a `signal-to-noise ratio' $\geq 35$ (see
Appendix). We confirmed that changing the color$\times$magnitude bin
sizes from 0.2$\times$0.3 mag$^2$ to 0.4$\times$0.5 mag$^2$ yields
similar results. The cleaned stellar catalog contains 17,346 stars
with $V \le 27.0$ mag.

\section{Main Results}

\begin{figure}[ht!]
\plotone{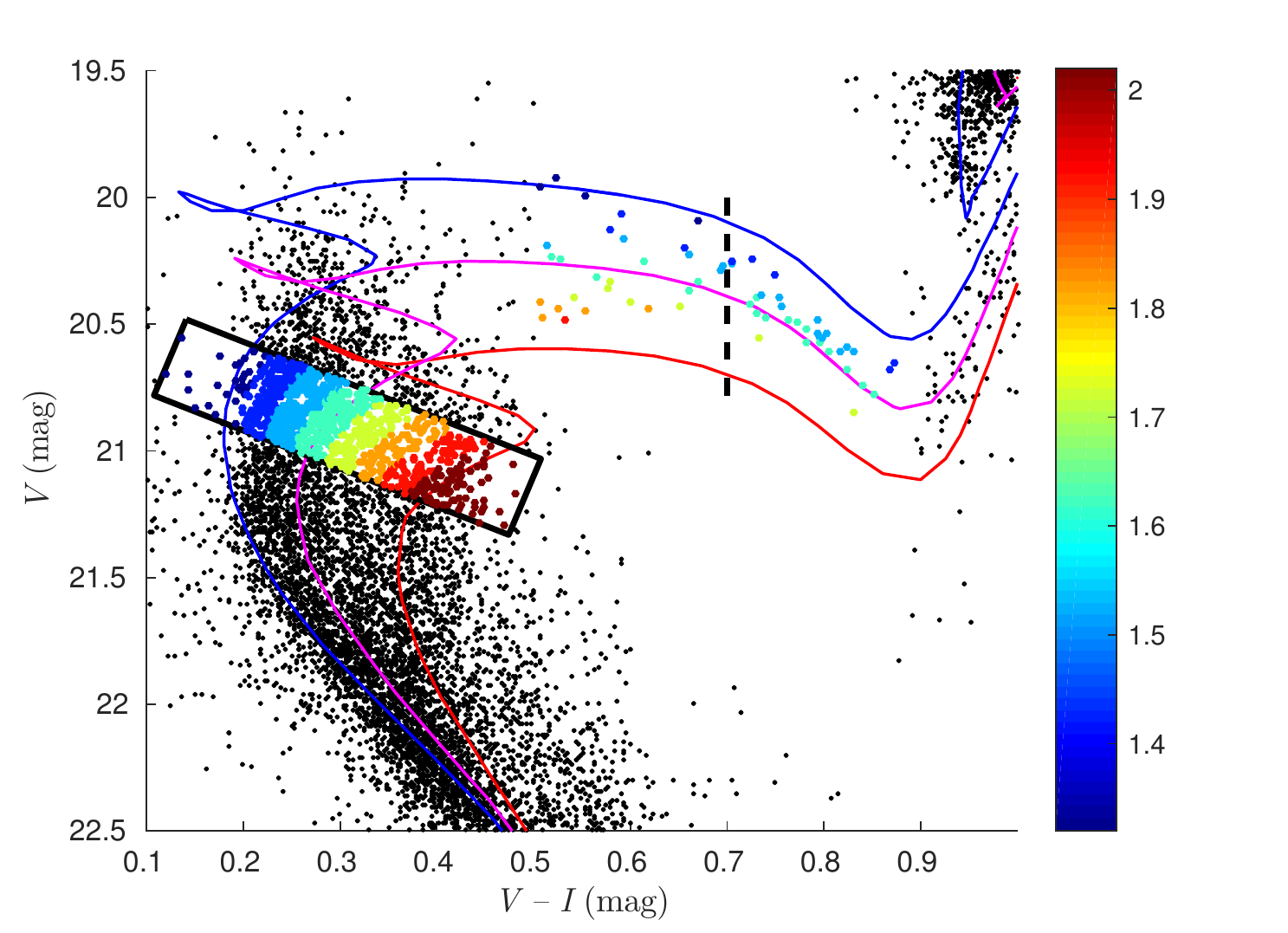}
\caption{NGC 419 CMD, showing isochrones of 1.32 Gyr (blue), 1.52 Gyr
  (magenta; corresponding to the peak in the cluster's SFH based on
  its red SGB morphology), and 2.02 Gyr (red). Colored points are the
  SGB stars and the eMSTO stars used to derive the SFHs (black box),
  assuming that these features can be described solely by stellar age
  ranges (indicated by the color bar; in Gyr). The vertical dashed
  black line shows where we separate the SGB stars into blue and red
  samples.}
\label{fig1}
\end{figure}

Figure 1 shows the NGC 419 CMD after field-star decontamination. The
cluster has an eMSTO at $20 \le V\le 22$ mag. We adopted the Padova
stellar evolution models \citep[PARSEC CMD 2.7,
  v. 1.2S;][]{Bressan12},\footnote{http://stev.oapd.inaf.it/cgi-bin/cmd$\_$2.7}
for $Z = 0.004$ \citep{Gira09,Glatt08}, and fitted the eMSTO extremes
with isochrones of $\log(t \mbox{ yr}^{-1})=9.12$ (1.32 Gyr; blue
line) and $\log(t \mbox{ yr}^{-1})=9.31$ (2.02 Gyr; red line),
adopting a distance modulus, $(m-M)_0 = 18.90$ mag and a visual
extinction $A_V = 0.181$ mag \citep{Rube10}. The maximum possible age
spread implied by the extent of the eMSTO is $\sim$700 Myr, similar to
the results of \citet[][$\sim$700 Myr]{Rube10} and \citet[][$\sim$670
  Myr]{Gira13}. However, the cluster exhibits a SGB that is broad on
the blue side and which becomes significantly narrower on the red
side. Previous studies of intermediate-age star clusters with eMSTOs
usually showed tight SGBs throughout \citep{Mack07,Li14b,BN15}
\citep[but see][]{Goud15}. This feature is, however, already apparent
in the NGC 419 data before field-star decontamination and is, hence,
not caused by our data reduction. In addition, the significance level
of our field-star decontamination is very high along the SGB, while
the good agreement between our {\sc iraf/daophot} and {\sc dolphot}
stellar catalogs further confirms the reality of the SGB
morphology. Examination of the SGB stars' spatial distribution reveals
that blending is unlikely responsible for the observed narrowing
either. The NGC 419 CMD of \citet{Goud14} shows a similar trend,
although that of \citet{Glatt08} resembles an SSP more closely. The
latter CMD is, however, composed of {\it HST} ACS/High Resolution
Channel observations covering a small stellar sample located in the
cluster's central region only. Our data are consistent with the
\citet{Glatt08} CMD for stars drawn from inside the cluster's core
radius. We thus conclude that the observed narrowing of the NGC 419
SGB is real and not caused by artifacts associated with our data
reduction (see Appendix).

To characterize the narrowing of the SGB, we derived SFHs based on the
morphologies of the blue and red sides of the SGB. We first selected
SGB stars at $0.50 \le (V-I) \le 0.89$ mag, where the blue boundary is
the red boundary of the eMSTO (determined by the oldest isochrone
used), while at the red boundary the model SGB luminosity reaches a
minimum. We divided the full sample into blue and red subsamples at
$(V-I) = 0.7$ mag. The blue SGB stars span a broad range in
luminosity, while on the red side almost all SGB stars converge to a
significantly narrower distribution. For each SGB star we determined
the best-fitting isochrone, adopting isochrones from 1.32 Gyr to 2.02
Gyr, in steps of 100 Myr. We counted the fractions of SGB stars
pertaining to each isochrone to derive the SGB's SFHs. The SGB
lifetime is a monotonically increasing function of the isochrone age
adopted, ranging from 9 Myr for the 1.32 Gyr isochrone to 22 Myr for
the 2.02 Gyr isochrone \citep{Bressan12}, i.e., significantly shorter
than our age resolution of 100 Myr. The small range in SGB lifetimes
therefore does not affect the SFHs derived. In addition, the
probability of an SGB star to be associated with a given isochrone is
nearly the same for all ages.

Typical uncertainties in our SGB photometry are very small ($\Delta V
\leq\ 0.040$ mag, $\Delta I \leq\ 0.044$ mag), corresponding to a $\la
50$ Myr age uncertainty. The isochrone fits are therefore
well-constrained. We averaged the SFHs derived from the catalogs of
100 field-star decontamination runs to minimize the effects of
randomly removing stars. The average numbers of blue and red SGB stars
are both 31. Figure 2 shows the SFHs derived for the blue and red SGB
subsamples. The former is clearly much flatter and more extended than
the latter, which is characterized by a sharp peak at 1.62 Gyr (see
the magenta isochrone in Fig. 1) and falls off steeply on either
side. The ratios of the periods stars spend in the red versus blue SGB
regions are approximately the same for all adopted ages, $\sim$1.3
\citep{Bressan12}. Hence, our derived SFHs reliably represent the real
SFHs. Note that while the apparently broadened base of the red-giant
branch (RGB) in the CMD is affected by lingering contamination by
field RGB stars owing to statistical fluctuations in the small numbers
of RGB cluster and field stars, field contamination does not affect
the morphology of the SGB to the same extent.

\begin{figure}[ht!]
\plotone{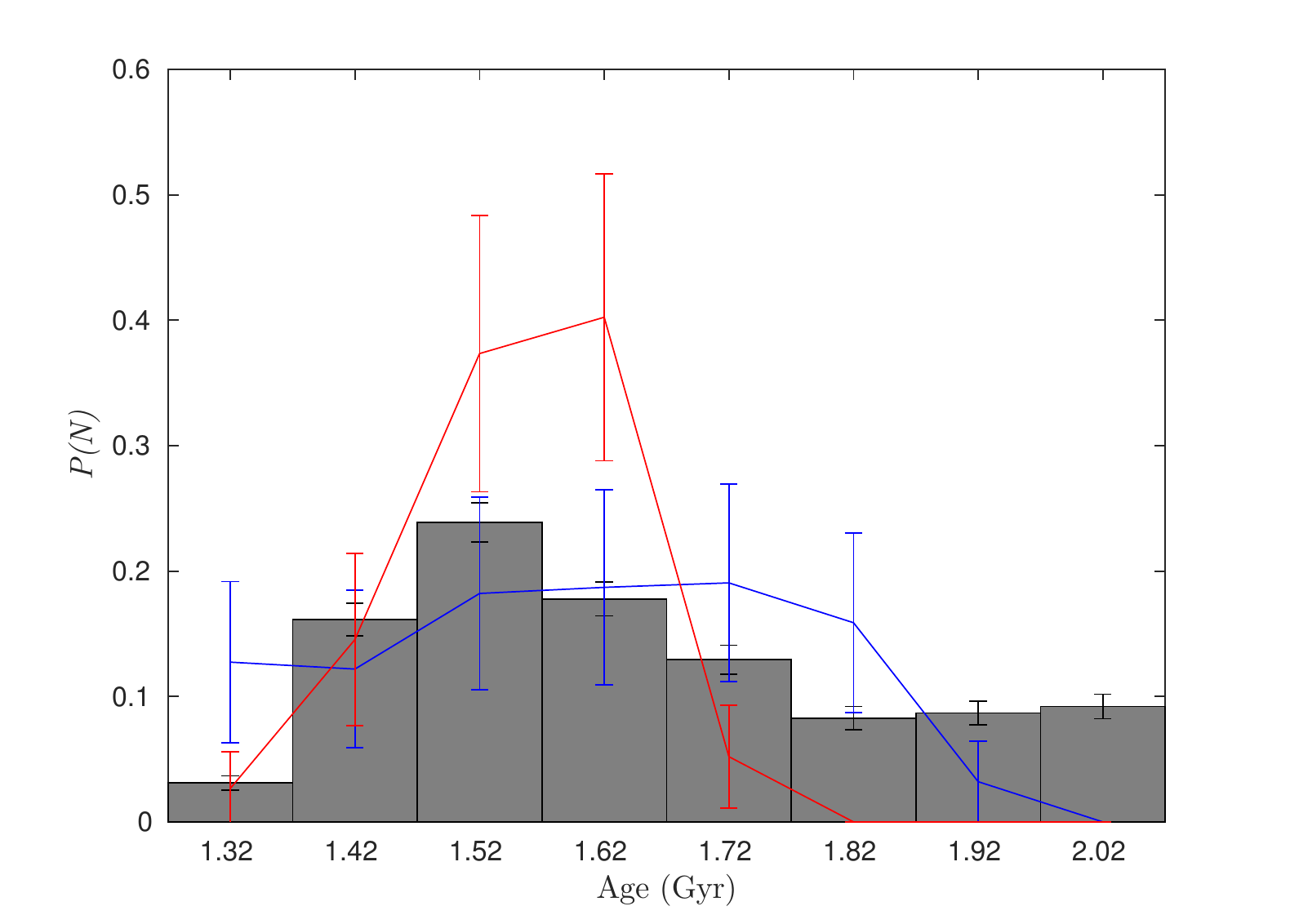}
\caption{SFHs derived from our sample SGB and eMSTO stars. Blue, red
  lines: SFHs of the blue, red sample SGB stars, respectively. Gray
  histogram: SFH of the eMSTO stars. The error bars represent
  1$\sigma$ Poissonian uncertainties.}
\label{fig2}
\end{figure}

We next derived the most representative eMSTO SFH. We first selected a
representative eMSTO sample, shown as the black box in Fig. 1. Its
left- and right-hand boundaries run approximately parallel to the
isochrones, while the top and bottom boundaries are oriented
perpendicularly to the isochrones. Our selection was made such that
the isochrones are well separated from each other, while the influence
of binary stars is minimized \citep[e.g.,][]{Goud14}. For each of our
967 eMSTO stars, we determined the best-fitting isochrone, using the
same isochrone set as for our SGB analysis. Figure 2 shows the SFH
derived for the eMSTO (gray histogram), indicating an age spread from
1.32 Gyr to 2.02 Gyr. In the presence of small-number statistical
fluctuations, this may be consistent with the SFH resulting from the
blue part of the SGB region, but certainly not with its red SFH. We
constructed 1000 synthetic clusters characterized by the eMSTO's SFH
to check whether the lack of old red SGB stars could have been caused
by stochastic effects. The results imply that the `convergence' of the
NGC 419 SGB is unlikely caused by stochastic sampling; for instance,
the probability of detecting fewer than 10\% of the red SGB stars near
the oldest isochrones is $P < 0.05$. This hence raises significant
doubts as to the presence of a genuine age spread in the cluster. It
is thus essential to consider alternative explanations for the
observed variation in the NGC 419 SGB morphology.

\section{Discussion}

A significant number of recent papers have focused on the presence of
rapidly rotating stars as a possible explanation of the observed
eMSTOs \citep{Bast09,Yang13,Li14a,Li14b,Bran15,Nied15}. Stellar
rotation affects the eMSTO area in two main ways, i.e., through
gravity darkening and rotational mixing. Gravity darkening leads to
luminosity and surface-temperature reductions owing to the centrifugal
force, which enlarges the eMSTO area. Rotational mixing causes an
increase in the sizes of the stellar convective cores, which prolongs
the main-sequence lifetime and can counteract the effects of
rotational mixing on the SGB. However, \cite{Yang13} pointed out that
stellar rotation could explain the origin of eMSTOs if a moderate
rotational mixing efficiency is adopted \cite[see
  also][]{Bran15,Nied15}. \cite{Li14b} concluded that the tight SGB in
NGC 1651 could be explained by a population of rapidly rotating stars
if gravity darkening were dominant.

For NGC 419, we suspected that the converging trend along its SGB may
be explained by rotational deceleration. Conservation of angular
momentum implies that the expansion of SGB stars may eventually result
in identical rotational evolution of both rapidly and slowly rotating
stars. We used the Geneva stellar rotation models
\citep{Eskt12,Geor13}\footnote{http://obswww.unige.ch/Recherche/evol/-Database-}
to explore whether the deceleration of rapid rotators can plausibly
explain the SGB morphology of NGC 419.

We acquired four synthetic clusters, two characterized by a
metallicity of $Z=0.002$ and the other two for $Z=0.006$. For each
metallicity we modeled one cluster with a flat initial rotation
distribution and another with its initial rotation distribution based
on \cite{Huang10}. We explored the effects of stellar rotation for
B-type stars with nine different rotation rates, $\Omega_{\rm ini} /
\Omega_{\rm crit}$ = 0.0, 0.1, 0.2, 0.3, 0.4, 0.5, 0.6, 0.7, 0.8, 0.9,
and 0.95; $\Omega_{\rm ini}$ is the initial stellar surface angular
velocity and $\Omega_{\rm crit}$ is the critical or `break-up' angular
velocity \citep{Geor13}. All synthetic clusters have the same age,
$\log(t \mbox{ yr}^{-1})=9.1$, i.e., the oldest available age for
B-type stars (with masses $m_\ast \ge 1.7 M_\odot$) and the closest
model age to that of NGC 419.

Figure 3 (left) shows the MSTO and SGB CMD regions for the two
synthetic clusters with flat initial rotation distributions; the color
bar represents $\Omega / \Omega_{\rm crit}$ at $\log(t \mbox{
  yr}^{-1})=9.1$. We adopted the same $A_V$ and distance modulus as
for NGC 419 and implemented realistic photometric uncertainties. Fits
of non-rotating Geneva isochrones to these eMSTOs imply age spreads of
$\sim$630 Myr and $\ge 460$ Myr for $Z=0.002$ and $Z=0.006$,
respectively.\footnote{The lack of SGB stars along the non-rotating
  isochrone of $log\log(t \mbox{ yr}^{-1})=9.1$ in the $Z=0.002$ CMD
  is caused by the B-type stars with $\Omega_{\rm ini} / \Omega_{\rm
    crit}\leq 0.3$ already having left the SGB.} Although the
left-hand panels of Fig. 3 seem to suggest that none of the SGBs in
our synthetic clusters converge into narrow sequences on the red side,
it appears that the rapid rotators might converge into a narrower
sequence, while the slow rotators occupy a much broader luminosity
range.

\begin{figure}[ht!]
\plotone{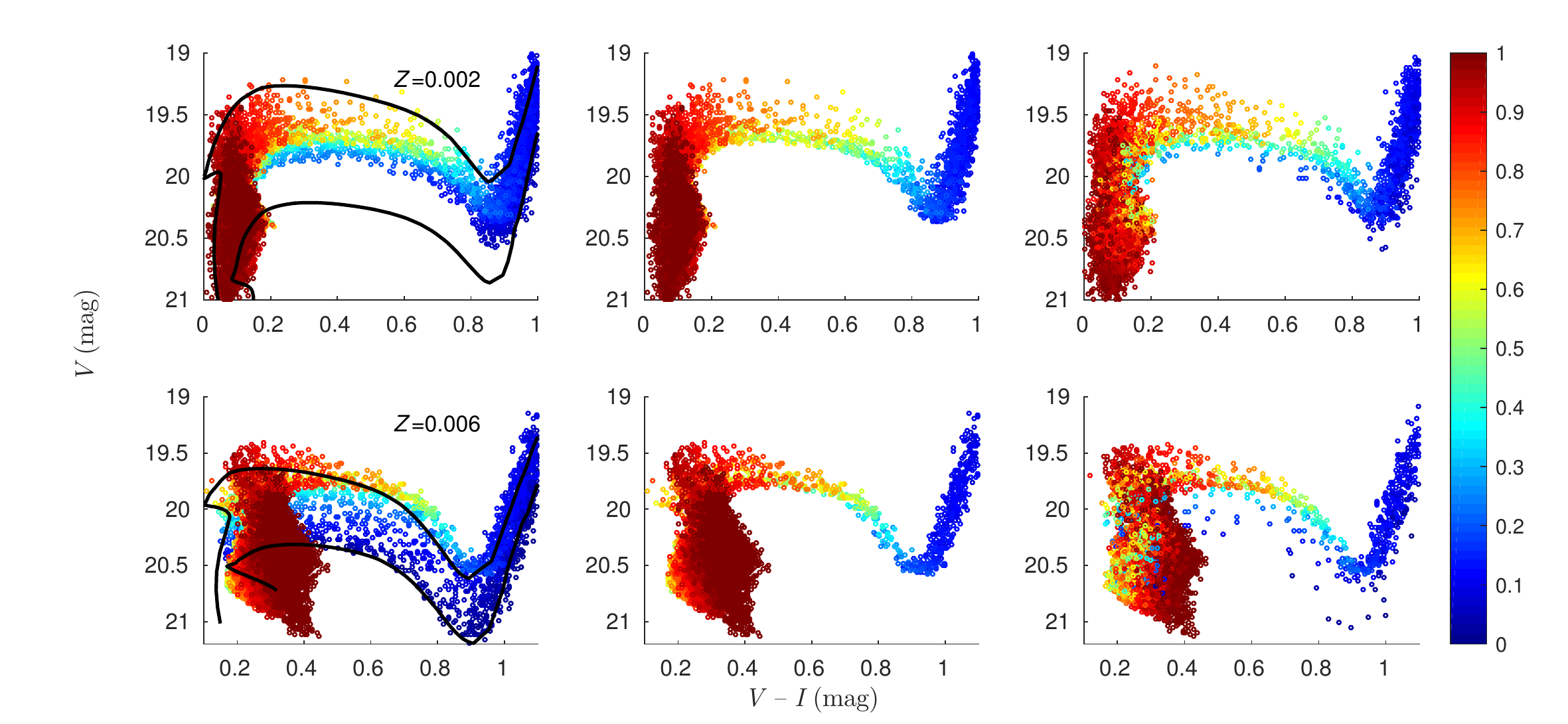}
\caption{Synthetic cluster CMDs based on the Geneva stellar rotation
  models for (top) $Z=0.002$ and (bottom) $Z=0.006$. Left: Adopting a
  flat initial rotation distribution, $\Omega_{\rm ini} / \Omega_{\rm
    crit} \in [0,1]$. Color bar: $\Omega / \Omega_{\rm crit}$ at
  $\log(t \mbox{ yr}^{-1})=9.1$. Black lines: $Z=0.002$, non-rotating
  isochrones for $\log(t \mbox{ yr}^{-1})=8.8$ and $\log(t \mbox{
    yr}^{-1})=9.1$; $Z=0.006$, non-rotating isochrones for $\log(t
  \mbox{ yr}^{-1})=8.9$ and $\log(t \mbox{ yr}^{-1})=9.1$. Middle: As
  the left panels, but for a minimum at $\Omega_{\rm ini} /
  \Omega_{\rm crit} = 0.5$. Right: As the left panels, but for the
  initial rotation distribution of \cite{Huang10}. Note that the $V$
  and $I$ magnitudes in the Geneva models refer to the
  Johnson--Cousins photometric system.}
\label{fig3}
\end{figure}

Hence we implemented a series of cut-offs to the initial rotation
rates to explore the effects of varying the minimum $\Omega_{\rm ini}
/ \Omega_{\rm crit}$ rate on the SGB morphology. \cite{Huang10}
determined that most low-mass ($2<m_\ast/M_{\odot}<4$) B-type stars
are born as rapid rotators: the fraction of their slowly rotating
sample stars, with rotation-velocity ratios $V_{\rm eq}/V_{\rm
  crit}<0.5$, is 37\%. We found that the convergence of the SGB is
most evident for initially rapid rotators: see Fig. 3 (middle), where
the CMDs are composed of stars with $\Omega_{\rm ini} / \Omega_{\rm
  crit}> 0.5$. The color bar represents $\Omega / \Omega_{\rm crit}$
at $\log(t \mbox{ yr}^{-1})=9.1$. Assessment of the colored data
points also shows that the SGB stars on the red side of the CMDs have
all slowed down to approximately the same $\Omega / \Omega_{\rm crit}
\le 0.3$--0.4 at $\log(t \mbox{ yr}^{-1})=9.1$.

The eMSTO stars in NGC 419 may be rotating faster than expected, given
that the converging cluster SGB suggests that gravity darkening may
indeed be important. Metallicity variations may also have a
significant impact on stellar atmospheric convection. When rapid
rotators are compared with slow rotators, rotational mixing dominates
the SGB morphology, because slowly rotating stars occupy the bottom
part of the SGB and broaden it to a large extent. However, the
assumption of a cluster to contain only rapid rotators seems
unrealistic. Nevertheless, the young clusters NGC 1856 \citep{DAnt15}
and NGC 1755 \citep{Milo16} both appear to consist of two-thirds of
rapid rotators. Therefore, we explored synthetic CMDs based on the
initial rotation distribution of \cite{Huang10}: see Fig. 3
(right). Note that at $\log(t \mbox{ yr}^{-1})=9.1$, all stars in the
$Z=0.002$ CMD with $\Omega_{\rm ini} / \Omega_{\rm crit} \leq 0.3$
will already have evolved off the SGB. However, the $Z=0.006$ cluster
clearly shows that the initially rapid rotators outnumber their
(initially) slowly rotating counterparts. The probability of finding
initially rapid rotators along real cluster SGBs is, therefore, much
higher than that for slow rotators. In addition, initially slowly
rotating SGB stars in the $Z=0.006$ CMD are faint. If such stars are
present in the NGC 419 CMD, they may be mixed with field stars and,
hence, will have been removed. It is therefore likely that NGC 419 is
indeed composed of a large fraction of rapid rotators. Unfortunately,
it is still cumbersome to model synthetic clusters with metallicities
other than $Z=0.002, Z=0.006$, and $Z=0.014$, or ages $\log(t \mbox{
  yr}^{-1})>9.1$. Therefore, the effects of stellar rotation on the
SGB morphology of real star clusters are still tentative. Published
CMDs also suggest that NGC 1852 \citep{Goud14}, NGC 2154
\citep{Goud14,Milo09}, NGC 2203 \citep{Goud14}, and Hodge 7
\citep{Milo09} may exhibit similar converging trends along their SGBs,
although somewhat less obvious than that in NGC 419.

\section{Summary and Conclusions}

We obtained PSF photometry from archival {\sl HST} images of the Small
Magellanic Cloud star cluster NGC 419 using two independent software
packages, {\sc iraf/daophot} and {\sc dolphot}. The CMDs resulting
from both stellar catalogs are mutually consistent and show a
narrowing trend along the cluster's SGB, i.e., the NGC 419 SGB appears
to be significantly narrower at redder than at bluer colors.

Initially assuming, for simplicity and in line with previous work,
that the widths of our CMD features may be driven entirely by stellar
age distributions, we derived SFHs using a sample of eMSTO stars, as
well as stars on the blue and red sides of the cluster's SGB. The SFH
for the red SGB exhibits a sharp peak at an age of 1.52 Gyr and drops
off steeply on either side of the peak, while the blue SGB morphology
is consistent with a much larger age range, from 1.32 Gyr to 1.92
Gyr. Moreover, the SFH of the eMSTO stars suggests an even broader age
range, from 1.32 Gyr to 2.02 Gyr. These three independently derived
SFHs challenge the postulated presence of a significant age spread to
explain the observed eMSTO extent. Hence, we considered the presence
of a population of stars with different stellar rotation rates.

For a synthetic cluster of age $\log(t \mbox{ yr}^{-1})=9.1$ and
$Z=0.002$ or $Z=0.006$, the SGB's convergence is most prominent for
stars with $\Omega_{\rm ini} / \Omega_{\rm crit} > 0.5$, which
suggests that NGC 419 may be composed of a large fraction of rapidly
rotating stars. Future improvements of stellar rotation models are
urgently needed to reach robust conclusions as to the importance of
variations in stellar rotation rates in intermediate-age star
clusters.

\section*{Acknowledgements}

We thank Yi Hu for his assistance with the {\sc dolphot} photometry
and with processing of the stellar catalogs. X. H. W. is grateful for
support from Peking University's Junzheng Fund for Undergraduate
Research. C. L. is partially supported by Strategic Priority Program
`The Emergence of Cosmological Structures' of the Chinese Academy of
Sciences (grant XDB09000000) and by a Macquarie Research
Fellowship. R. d. G. and L. D. acknowledge research support from the
National Natural Science Foundation through grants 11073001, 11373010,
and 11473037.

\appendix

\setcounter{figure}{3}

Here we introduce the details of our data reduction, including the
determination of the cluster center, its radius, and the procedures
used for field-star decontamination. We also summarize the matching,
cleaning, and comparison of the {\sc dolphot} and {\sc iraf/daophot}
stellar catalogs.

\bigskip
\subsection*{{\sc dolphot} and {\sc iraf/daophot} Photometry}

The {\sc dolphot} software generated one stellar catalog for each pair
of images taken in the $V$ and $I$ bands, so that we have four initial
stellar catalogs pertaining to the deep exposures and two stellar
catalogs resulting from the shallower exposures. `Bad' sources were
removed following the prescriptions in the {\sc dolphot} manual
\citep{Dolphin13}. We also referred to \cite{Mack07}'s data reduction
of NGC 1846 when removing the `bad' objects. The cleaned initial
stellar catalogs only contain stars with $-0.15 \leq$ sharpness $\leq$
0.15, crowding $\leq$ 0.5, and photometric errors $\leq$ 0.2 mag. We
also only retained sources with object type 1, i.e., stars classified
by {\sc dolphot} as `good.' Next, we converted the stars' $x$ and $y$
positions on the images into right ascension and declination, using
the python package {\sc drizzlepac}, which corrects for astrometric
image distortions. Using the stellar positions on the sky as the
criterion for matching stars in different stellar catalogs, we
combined the four long-exposure stellar catalogs into a single
catalog, where we only selected those stars that appear in all four
images. The stellar catalogs resulting from the shorter exposure times
were similarly combined. Subsequently, we combined the deep and
shallow stellar catalogs, since bright stars are saturated in the long
exposures and faint stars cannot be detected in the short
exposures. If a star appeared both in the long- and short-exposure
catalogs, we included the photometry from the deeper observation. The
same procedures were used to obtain the stellar catalog pertaining to
the nearby background field.

As regards the {\sc iraf/daophot} photometry, the four long-exposure
images were first aligned and combined into a single image. PSF
photometry was then performed on the combined image, resulting in a
single stellar catalog containing all $V$-band information and a
second catalog containing the $I$-band photometry. We then applied a
series of parameters to clean the two stellar catalogs, including
cut-offs in the sharpness and in the number of iterations ($n_{\rm
  iter}$) it took to properly fit a star. Next, we matched the $V$-
and $I$-band catalogs to assess the quality of the CMD. The final
parameters adopted were sharpness $\in [-0.3,0.3]$, $n_{\rm iter} <
40$, and photometric errors $<$ 0.2 mag. These ranges produced a CMD
that is sufficiently `clean' and a stellar catalog that is
sufficiently complete for the purposes of this paper; limiting the
sharpness value to a smaller range would remove too many
stars.\footnote{We also checked by comparison with the {\sc dolphot}
  stellar catalog what the effect would be of changing the range in
  sharpness value from $[-0.15,0.15]$ to $[-0.3, 0.3]$. This did not
  affect the SGB morphology.}

\subsection*{Artificial Star Tests}

We performed artificial star tests using {\sc dolphot} to assess the
spatially dependent completeness levels. We generated $\sim$750,000
fake stars, which we added to the raw images, and performed PSF
photometry on these stars. To ensure that the fake stars did not
significantly increase the crowding of the images, we only added 100
fake stars to an image at any time. After all artificial stars had
been measured properly, we removed `bad' sources using the same
criteria as adopted for removing `bad' sources from the catalog of
real cluster stars. Stars whose measured magnitudes differed more than
0.25 mag from their input magnitudes and those whose $(x,y)$ positions
on the images differed by more than 0.5 pixels from the input
positions were considered to not have been recovered. Using these
criteria, we obtained a catalog of `recovered' fake stars. We next
calculated the distance of each fake star to the cluster center (see
below) and divided them into radial and magnitude bins. We used these
bins, in turn, to calculate the fractions of the number of `recovered'
stars with respect to the original number of stars. These fractions
represent the spatially dependent completeness levels, which we used
in the determination of the cluster's structural parameters. The full
two-dimensional spatially dependent completeness surface is shown in
Fig. 4 (left).

\begin{figure}[ht!]
\includegraphics[angle=90,width=18cm]{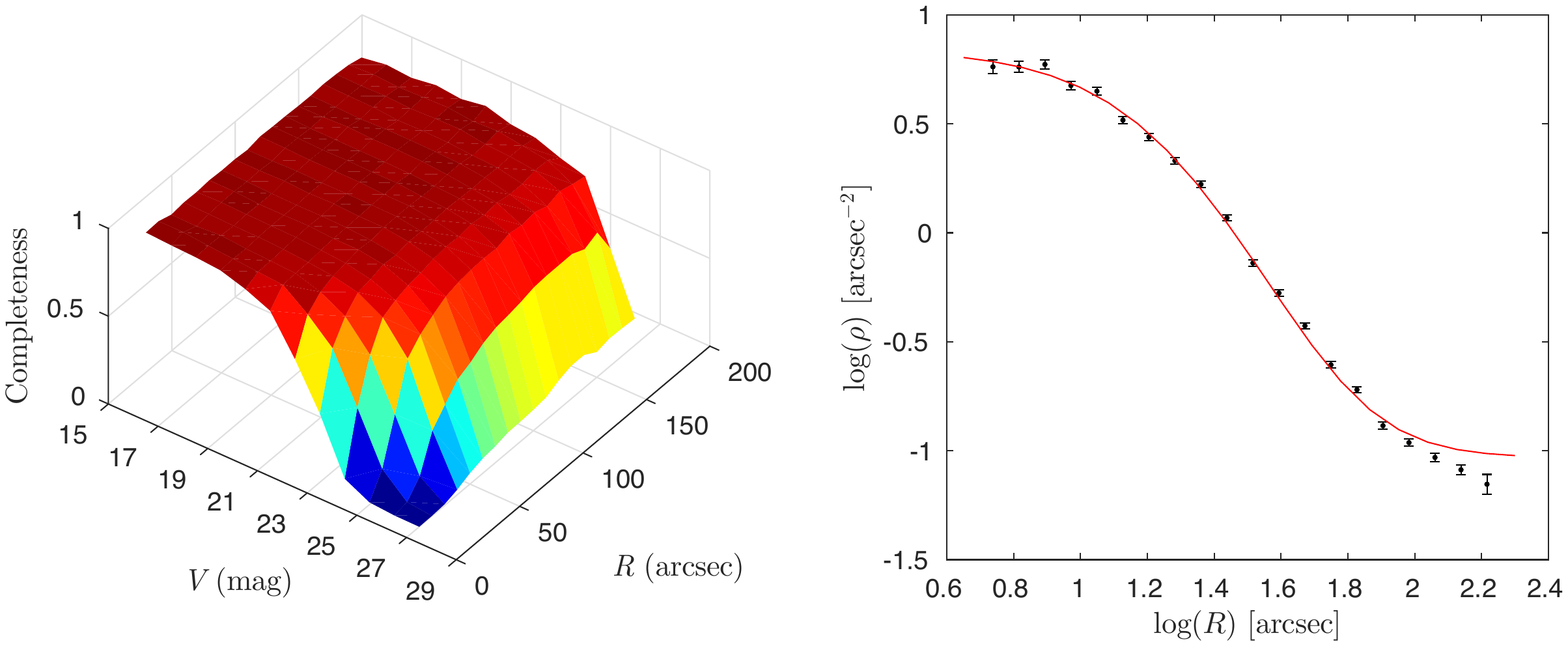}
\caption{Left: Spatially dependent completeness levels as a function
  of cluster radius and stellar magnitude. Right:
  Completeness-corrected radial number-density profile of NGC 419.
  The 1$\sigma$ uncertainties are defined by Poissonian noise
  statistics. The red curve is the best-fitting EFF profile.}
\label{fig4}
\end{figure}

\subsection*{Cluster Structure}

We determined the cluster's structural parameters by calculating the
(projected) radial stellar number-density profile. To do so, we
counted the numbers of stars in different rings, $N(R)$, and divided
them by the areas of the rings, $A(R)$. The number of stars in each
ring was corrected for the spatially dependent completeness. However,
since the completeness level near the cluster center is lower than
that in the outer regions, faint stars that can be detected in the
outer parts cannot always be detected in the inner regions. Therefore,
we only used stars brighter than $V = 22.5$ mag when calculating the
cluster's number-density profile, because at this brightness the
completeness curve of the innermost ring decreases to 50\%.

The areas of the rings were estimated using a Monte Carlo-type method,
because some of the rings are not completely covered by the
observations. Specifically, we first calculated the total area of the
region observed and then generated a million points, homogeneously
distributed across the entire region. We then used the number fraction
of points located in each ring, multiplied by the total area of the
full region to represent the area of each ring. The radial
number-density profile, $f(R)$, is then obtained, $f(R)=N(R)/A(R)$. We
repeated this process six times and averaged the resulting radial
number-density profile to minimize the random errors introduced by the
Monte Carlo method. The average radial number-density profile is shown
in the right-hand panel of Fig. 4, where the data points are
represented by black points. The error bars represent 1$\sigma$
Poissonian uncertainties.

Since NGC 419 is an intermediate-age star cluster, we used an EFF
profile to fit its radial number-density profile,
\begin{equation}
f(r)=f_0 (1+r^2/a^2)^{-\gamma /2}+\phi ,
\end{equation}
where $f_0$ is the central projected stellar density, $a$ is a measure
of the core radius, and $\gamma$ is the power-law slope. The parameter
$\phi$ is added to represent background contamination. The results of
our fits are $f_0=8.88$ arcsec$^{-2}$, $a=19.17''$, $\gamma=3.43$, and
$\phi=0.093$ arcsec$^{-2}$. The radius where the stellar number
density drops to half the central value is the core radius: $r_{\rm
  c}=15''$. The cluster radius is defined as five times the core
radius: $R$=$75''$. This radius is sufficiently large to select a
relatively complete sample of SGB stars; adopting a cluster radius
that is much larger will cause considerable contamination of the SGB
region by field stars.

\subsection*{SGB Morphology Confirmation}

\subsubsection*{SGB Morphology Comparison}

The left-hand and middle panels of Fig. 5 show the CMDs of the cluster
stars at $R<75''$ based on the {\sc dolphot} and {\sc iraf/daophot}
catalogs, respectively. The converging trend along the SGB is indeed
clearly visible prior to us having performed field-star
decontamination. To confirm this trend, we first made comparsions
between our two stellar catalogs. We show zoom-ins of the SGB regions
in the CMDs of both catalogs in the insets of the left-hand and middel
panels of Fig. 5, for stars located at $R \le 30''$. We used a radius
of twice the core radius for this check, because the number of SGB
stars inside $R_{\rm core} = 15''$ is too small, while the converging
trend also disappears for stars located inside the core radius. On the
other hand, a radius that is too large can cause significant
contamination by field stars. The SGB stars, with $0.5 \le (V-I)\le
0.89$ mag, are shown as magenta squares in those insets of Fig. 5. We
matched the two samples of SGB stars and averaged their magnitudes to
quantify the effects of averaging on the SGB morphology.

Among the 54 SGB stars in the {\sc iraf/daophot} catalog and the 48
SGB stars in the {\sc dolphot} catalog, 38 stars were matched. The
`average' SGB for $R \le 30''$ is shown in the right-hand panel of
Fig. 5, where the orange and green squares are the matched SGB and the
magnitude-averaged SGB stars, respectively. It is clear that the
converging trend exhibited by the SGB stars in the {\sc dolphot}
stellar catalog does not change significantly, which strongly suggests
that this feature is intrinsic and not an artifact introduced by our
photometric analysis or our method of matching stars.

\begin{figure}[ht!]
\plotone{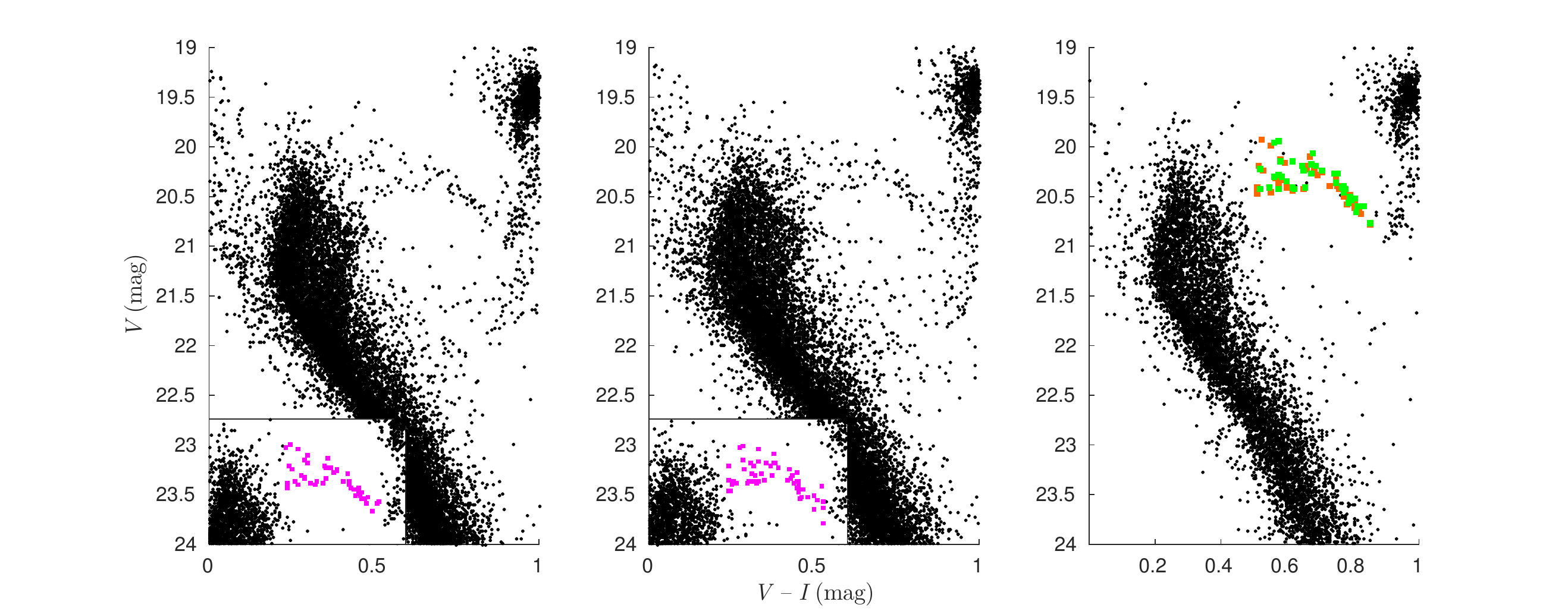}
\caption{CMDs of NGC 419. Left, middle: based on the {\sc dolphot},
  {\sc iraf/daophot} stellar catalogs for stars at $R \le
  75''$. Insets: SGB zoom-ins for stars within $R = 30''$. Magenta
  squares: SGB stars. Right: CMD of the stars within $R = 30''$ from
  the {\sc dolphot} catalog. Orange squares: matched SGB stars from
  the {\sc dolphot} catalog. Green squares: magnitude-averaged,
  matched SGB stars.}
\label{fig5}
\end{figure}

\subsection*{Field-Star Decontamination}

Since the number of stars inside $R=75''$ is significantly affected by
the background stellar population, statistical decontamination of the
background field must be done. In this section we demonstrate that the
observed SGB morphology is not caused by our approach to field-star
decontamination. The method adopted for field-star decontamination is
similar to that employed by \cite{Li14b}. We first used a Monte Carlo
method to estimate the area of the region inside $R=75''$. The
fraction of the area inside this radius divided by the total area of
the frame is 0.425.

We next counted the number of field stars in each color--magnitude bin
and randomly deleted 0.425 times that number of stars from the
corresponding color--magnitude bin of the CMD of the star cluster. If
the number of stars that had to be removed from a bin was larger than
the number of stars in the corresponding bin of the CMD of the
cluster, we removed all stars in such a bin. In addition, we used a
similar method to that adopted by \cite{Cabr16} to estimate the
significance of our field-star decontamination, where the
signal-to-noise ratio is defined as the number of stars remaining in
each color--magnitude bin after field-star decontamination divided by
the average dispersion of the number of field stars in the
corresponding bin. The significance levels of the color--magnitude
bins adopted here are shown in Fig. 6 (left); the SGB region has been
highlighted using red boxes. It is clear that the SGB region is
characterized by a signal-to-noise ratio $>35$. Specifically, any
number from 0 to 5 stars must be removed from the 32 SGB stars in the
blue region of the SGB, $0.5 \le (V-I) \le 0.7$ mag, and 2 to 5 stars
should be removed from the 34 stars occupying the red part of the SGB,
$0.7 \le (V-I) \le 0.89$ mag. The number of stars to be removed is
much smaller, if not negligible, compared with the total number of
stars located in those color ranges. Therefore, the SGB morphology is
not affected significantly by our field-star decontamination. We
confirmed that the field-decontamination quality is consistent for
reasonable changes in bin sizes. Bin sizes that are too small will not
result in adequate numbers of stars to be removed from a bin, while
bin sizes that are too large will not allow us to distinguish the
features of the field CMD from those of the cluster CMD.

The result of our field-star decontamination is shown in the middle
and right-hand panels of Fig. 6. The top middle panel shows the CMD of
the stars inside $R=75''$ in the cluster field. On the top right, we
show the same cluster CMD, but after field-star decontamination. The
bottom middle panel is the CMD of the entire field region, and the
bottom right-hand panel is the CMD composed of stars that were
randomly removed from the cluster region.

\begin{figure}[ht!]
\plotone{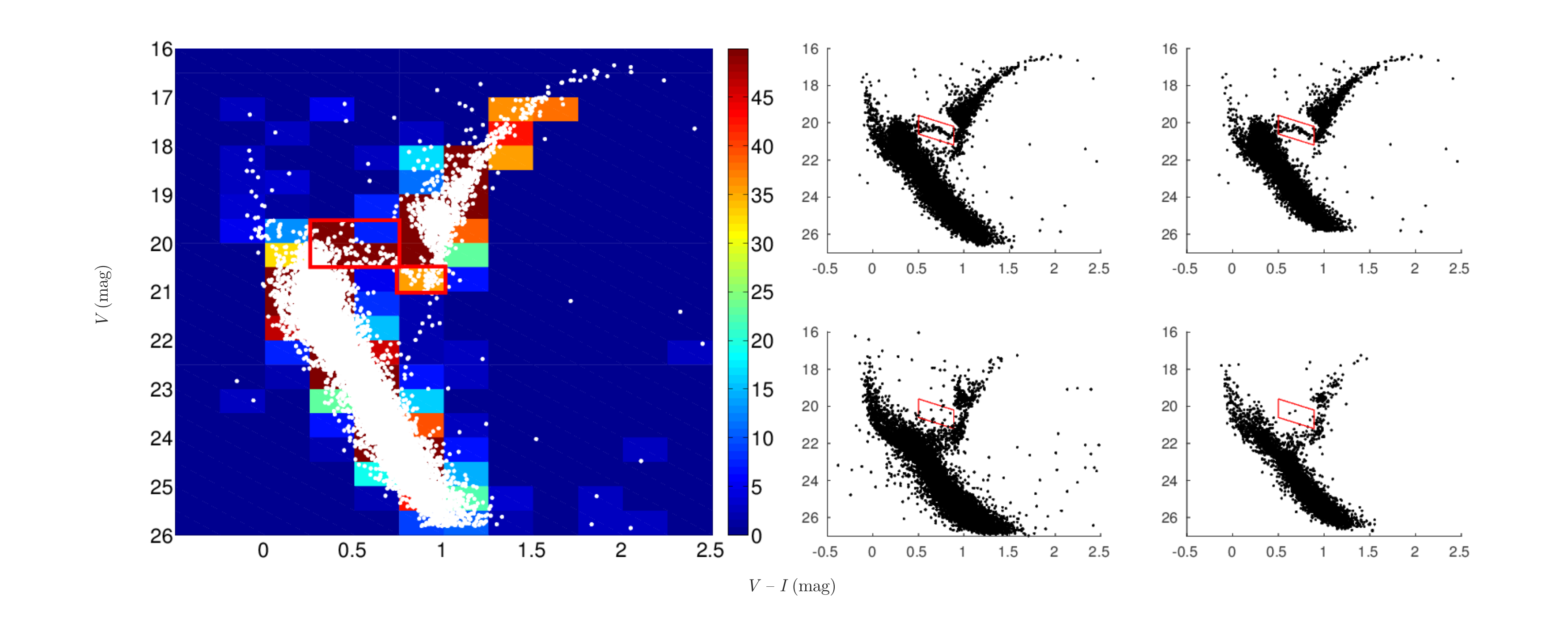}
\caption{Left: CMD of the stars located within $R=75''$ from the
  cluster center after field star decontamination. Different colors
  represent different signal-to-noise ratios. The SGB region is
  highlighted with red boxes. Right: Field-star decontamination. Top
  middle: CMD of the cluster stars ($R \le 75''$). Top right: CMD of
  the cluster after field-star decontamination. Bottom middle: CMD of
  the entire field region. Bottom right: CMD of the randomly removed
  stars in the cluster field by our field-star decontamination
  procedure.}
\label{fig6}
\end{figure}

\subsubsection*{Stochastic Variations}

We constructed synthetic clusters to estimate the probability of not
finding any old stars in the red part of the SGB because of stochastic
effects. Using the SFH of the eMSTO and our set of model isochrones,
we populated the main-sequence and SGB regions with stars randomly
drawn from the \citep{Kroupa01} stellar initial mass function. The
numbers of stars of the synthetic clusters were normalized to the
numbers of stars found in the main-sequence and SGB regions of the
observed cluster CMD. We added unresolved binary companions to 50\% of
the stars located on the main sequence (but none to the SGB stars),
based on the observational $\sim$20\% unresolved binary fraction, with
mass ratio $q\leq0.6$, assuming a flat mass-ratio distribution. We
obtained this unresolved binary fraction using the method of
\cite{Li14a}. It is close to that derived by \cite{Rube10}, who found
a binary fraction of $\sim$18\% for mass ratios greater than
0.7. Photometric errors were added to the stars, and the simulations
were run 1000 times. Each time we calculated two SFHs, for the blue
and red parts of the SGB. Figure 7 shows the results of our
simulations: the left-hand panel shows one typical synthetic cluster
characterized by an age spread based on the extent of the eMSTO, while
the right-hand panel illustrates the frequency of detecting a fraction
$P(N)$ of red SGB stars associated with the oldest isochrones, aged
1.82 Gyr, 1.92 Gyr, and 2.02 Gyr. The probability of not finding red
SGB stars near the oldest isochrones is very small indeed. For
instance, the frequency of finding fewer than 10\% of red SGB stars
near the oldest isochrones is $P < 0.05$. Therefore, stochastic
variations are not responsible for the converging trend seen in the
SGB of NGC 419.

\begin{figure}[ht!]
\plotone{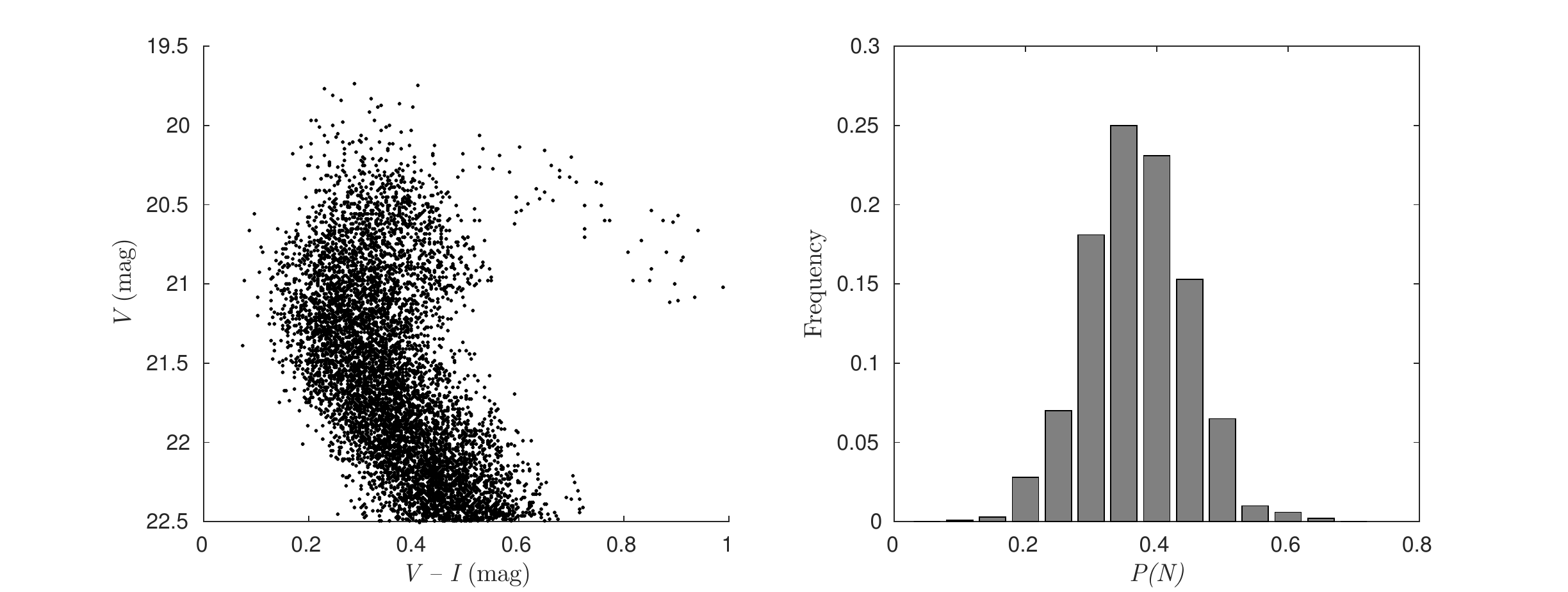}
\caption{Left: example of a synthetic cluster realization. Right:
  detection probability, $P(N)$, of red SGB stars to be associated
  with the oldest isochrones adopted (1.82 Gyr, 1.92 Gyr, and 2.02
  Gyr).}
\label{fig9}
\end{figure}

\end{document}